\documentclass[aps,prl,twocolumn,showpacs,superscriptaddress,preprintnumbers,amsmath,amssymb]{revtex4-1}

\usepackage{graphicx}
\usepackage{epsfig}
\usepackage{dcolumn}
\usepackage{bm}
\usepackage{longtable}
\usepackage{color}

\begin{document}


\title{CeCu$_2$Ge$_2$: Challenging our Understanding of Quantum Criticality}



\author{B.\ Zeng}
\affiliation{National High Magnetic Field Laboratory, Florida
State University, Tallahassee-FL 32310, USA}
\author{Q.\ R.\ Zhang}
\affiliation{National High Magnetic Field Laboratory, Florida
State University, Tallahassee-FL 32310, USA}
\author{D.\ Rhodes}
\affiliation{National High Magnetic Field Laboratory, Florida
State University, Tallahassee-FL 32310, USA}
\author{Y.\ Shimura}
\affiliation{Institute for Solid State Physics, University of Tokyo, Kashiwa 277-8581, Japan}
\affiliation{National High Magnetic Field Laboratory, Florida
State University, Tallahassee-FL 32310, USA}
\author{D.\ Watanabe}
\affiliation{Department of Physics, Kyoto University, Kyoto 606-8502, Japan}
\affiliation{National High Magnetic Field Laboratory, Florida
State University, Tallahassee-FL 32310, USA}
\author{R.\ E.\ Baumbach}
\affiliation{National High Magnetic Field Laboratory, Florida
State University, Tallahassee-FL 32310, USA}
\author{P. Schlottmann}
\affiliation{Department of Physics, Florida State University, Tallahassee, Florida 32306, USA}
\author{T.\ Ebihara}
\affiliation{Department of Physics, Graduate School of Science, Shizuoka University, Shizuoka 422-8529, Japan}
\author{L.\ Balicas} \email{balicas@magnet.fsu.edu}
\affiliation{National High Magnetic Field Laboratory, Florida
State University, Tallahassee-FL 32310, USA}


\date{\today}

\begin{abstract}
In the zero temperature limit, the zero-point quantum fluctuations of certain degrees of freedom (or quantum criticality) is claimed to describe the collective fluctuations of systems undergoing a second-order phase-transition. To date, some of the best studied examples of quantum phase-transitions, and concomitant anomalous physical behavior, involve $f-$electron magnetism in heavy-fermion metals, where quantum criticality (QC) is ascribed to either the suppression of a spin-density wave (SDW) ground-state or the Kondo-effect.
Here, we unveil evidence for a quantum phase-transition in CeCu$_2$Ge$_2$ which displays both an incommensurate spin-density wave (SDW) ground-state, and a strong renormalization of the quasiparticle effective masses ($\mu$) due to the Kondo-effect. For \emph{all} angles $\theta$ between an external magnetic field ($H$) and the crystallographic $c-$axis, the application of $H$ leads to the suppression of the SDW-state through a 2$^\text{nd}$-order phase-transition at a $\theta-$dependent critical-field $H_{p}(\theta)$ leading to the observation of small Fermi surfaces (FSs) in the paramagnetic (PM) state. For $H \| c$-axis, these FSs are characterized by light $\mu$s pointing also to the suppression of the Kondo-effect at $H_p$ with surprisingly, \emph{no} experimental evidence for quantum-criticality (QC). But as $H$ is rotated towards the \emph{a}-axis, these $\mu$s increase considerably becoming undetectable for $\theta > 56^{\circ}$ between $H$ and the \emph{c}-axis. Around $H_{p}^{a}\sim 30$ T the resistivity becomes $\propto T$ which, coupled to the divergence of $\mu$, indicates the existence of a field-induced QC-point at $H_{p}^{a}(T=0 \text{ K})$. This observation, suggesting FS hot-spots associated with the SDW nesting-vector, is at odds with current QC scenarios for which the continuous suppression of \emph{all} relevant energy scales at $H_{p}(\theta,T)$ should lead to a line of quantum-critical points in the $H-\theta$ plane. Finally, we show that the complexity of its magnetic phase-diagram(s) makes CeCu$_2$Ge$_2$ an ideal system to explore field-induced quantum tricritical and QC end-points.
\end{abstract}


\maketitle

\section{Introduction}

In the $T \rightarrow 0$ K limit, the \emph{continuous} suppression of an ordered-state,
leads to a singularity denominated quantum-critical point (QCP) which separates ordered and disordered states.\cite{si, senthil, gegenwart}
For intermetallic heavy-fermion compounds displaying an antiferromagnetic ground-state, two types of QCPs were identified:
i) the suppression to zero temperature of a Fermi surface (FS) instability which leads to a spin density wave (SDW) ground state
whose quantum-critical behavior is described in terms of the fluctuations of the SDW order parameter \cite{hertz,millis} and
ii)  the continuous destruction of the Kondo-effect at the AFM to paramagnetic transition,
which seems to lead to a discontinuous change in the Fermi surface volume across
the QCP.\cite{si, paschen, coleman, gegenwart2, friedemann} Here, we study the possibility of detecting magnetic field-induced quantum-criticality in CeCu$_2$Ge$_2$
since both, its antiferromagnetic SDW ground-state and the Kondo-effect, can be suppressed by an external magnetic-field.
If the suppression of either energy scale was continuous, the field would be expected to tune the system towards a quantum-critical point,
making CeCu$_2$Ge$_2$ an ideal system for probing the universality of either quantum-critical scenario.
Regardless of the field orientation, we find that both energy scales as well as the Kondo-coherence temperature are seemingly continuously suppressed by the
field. However, at low temperatures this does not necessarily lead to clear experimental evidence for quantum-criticality except when the field is applied along the $a$-axis.
This would be at odds with the above mentioned quantum-critical scenarios, for which the continuous suppression of all relevant energy scales should necessarily lead to a quantum-critical point, and therefore to a line of quantum-critical points in the field as a function of angle phase-diagram.

To date only few metallic systems such as YbRh$_2$Si$_2$, CeCoIn$_5$, CeRhIn$_5$ or Sr$_3$Ru$_2$O$_7$, have been claimed to display a field-induced quantum critical point.  Among those, even fewer are appropriate for the study of the Fermi surface evolution and associated quasi-particles across a field-induced QCP through the de Haas van Alphen effect. For example, in YbRh$_2$Si$_2$ the QCP occurs at fields too small for the observation of the dHvA-effect \cite{gegenwart,gegenwart2}. In cleaner Sr$_3$Ru$_2$O$_7$ single crystals the QCP is ``concealed" by the emergence of new electronic phase,\cite{grigera2} while in CeRhIn$_5$ the Fermi surface changes quite abruptly\cite{shishido} across a putative pressure-induced QCP\cite{park} which is difficult to understand within a continuous 2$^{\text{nd}}$-order phase-transition scenario. In CeCoIn$_5$ on the other hand, the the conventional Lifshitz-Kosevich formalism would seem to break down upon approaching a field-induced QCP with the renormalization of the quasiparticle effective mass being spin dependent.\cite{mcCollam} It is unclear at the moment if all quantum critical systems would display similar behavior. As we show below, CeCu$_2$Ge$_2$ offers us with the opportunity of exploring QC behavior in a SDW system characterized by strong Kondo-coupling through the de Haas van Alphen-effect and in the absence of the disorder inherent to alloys or the technical limitations imposed by high pressure measurements.

Both heavy-fermion intermetallics CeCu$_2$Ge$_2$ and CeCu$_2$Si$_2$ crystallize in the ThCr$_2$Si$_2$
crystallographic structure which is common to several Fe-pnictide superconductors and also to superconducting URu$_2$Si$_2$.
These compounds are characterized by Kondo ``coherence" temperatures, or the characteristic temperature below which a Kondo lattice
develops, ranging from 5 K (CeCu$_2$Ge$_2$) to $\sim 20$ K (CeCu$_2$Si$_2$).\cite{deboer,assmus}
According to terahertz spectroscopy, the development of a Kondo lattice in CeCu$_2$Ge$_2$ leads
to an enhancement, by one-order of magnitude, in the carrier effective-mass upon cooling from 3 K to values approaching $80$ $m_0$ at $\sim 1$ K, where $m_0$ is
the free-electron mass.\cite{armitage} Although CeCu$_2$Si$_2$ displays a superconducting ground-state, \cite{assmus, steglich} claimed to be unconventional and ``magnetically mediated" due to its proximity to an antiferromagnetic (AF) QCP (tuned by pressure or composition), \cite{stockert} CeCu$_2$Ge$_2$ develops long-range incommensurate AF-order below a N\'{e}el temperature $T_N \simeq 4.1$ K.\cite{krimmel} In the CeCu$_2$(Ge$_x$Si$_{1-x}$)$_2$ series $T_N$ is observed to decrease continuously from 4.1 K (for $x=1$) to $\lesssim 1$ K (for $x=0$).\cite{knebel} In CeCu$_2$Si$_2$ the type, or even the presence of AF-order (or superconductivity), is markedly dependent on the sample synthesis protocol. \cite{stockert, steglich2, faulhaber, arndt, stockert3} According to neutron scattering \cite{stockert2, stockert4, krimmel, dksingh} the AF-order propagation vector for both end compounds, and for concentrations in between, is incommensurate and close to \textbf{Q$_N$}$= (0.25,0.25,0.5)$ \cite{stockert4}, or more precisely \textbf{Q$_N$}$= (0.285,0.285,0.54)$ according to Ref. \onlinecite{dksingh}, and ascribed to FS nesting.\cite{stockert2, zwicknagl} This magnetic ground-state is determined by the splitting of the Ce $4f^1$ $(J=5/2)$ ground-state multiplet under the action of the tetragonal crystalline electric-field (CEF), which according to Ref. \onlinecite{loewenhaupt} leads to a ground-state doublet and an excited quasi-quartet consisting of two doublets at 17.0 and 18.3 meV, respectively. For either compound the application of pressure leads to the  stabilization of superconductivity around a critical-pressure $p_{c1}$ needed to fully suppress AF, \cite{jaccard, yuan} pointing to a pressure-tuned antiferromagnetic QCP. Nevertheless, the maximum superconducting transition temperature $T_c$ is observed around $p_{c2}( > p_{c1})$ where the unit cell volume collapses \cite{yuan, kobayashi, onodera}, an effect ascribed to an abrupt change in the valence of Ce suggesting perhaps a second QCP at $p_{c2}$.\cite{holmes, watanabe}
\begin{figure*}[htb]
\begin{center}
\includegraphics[width = 12 cm]{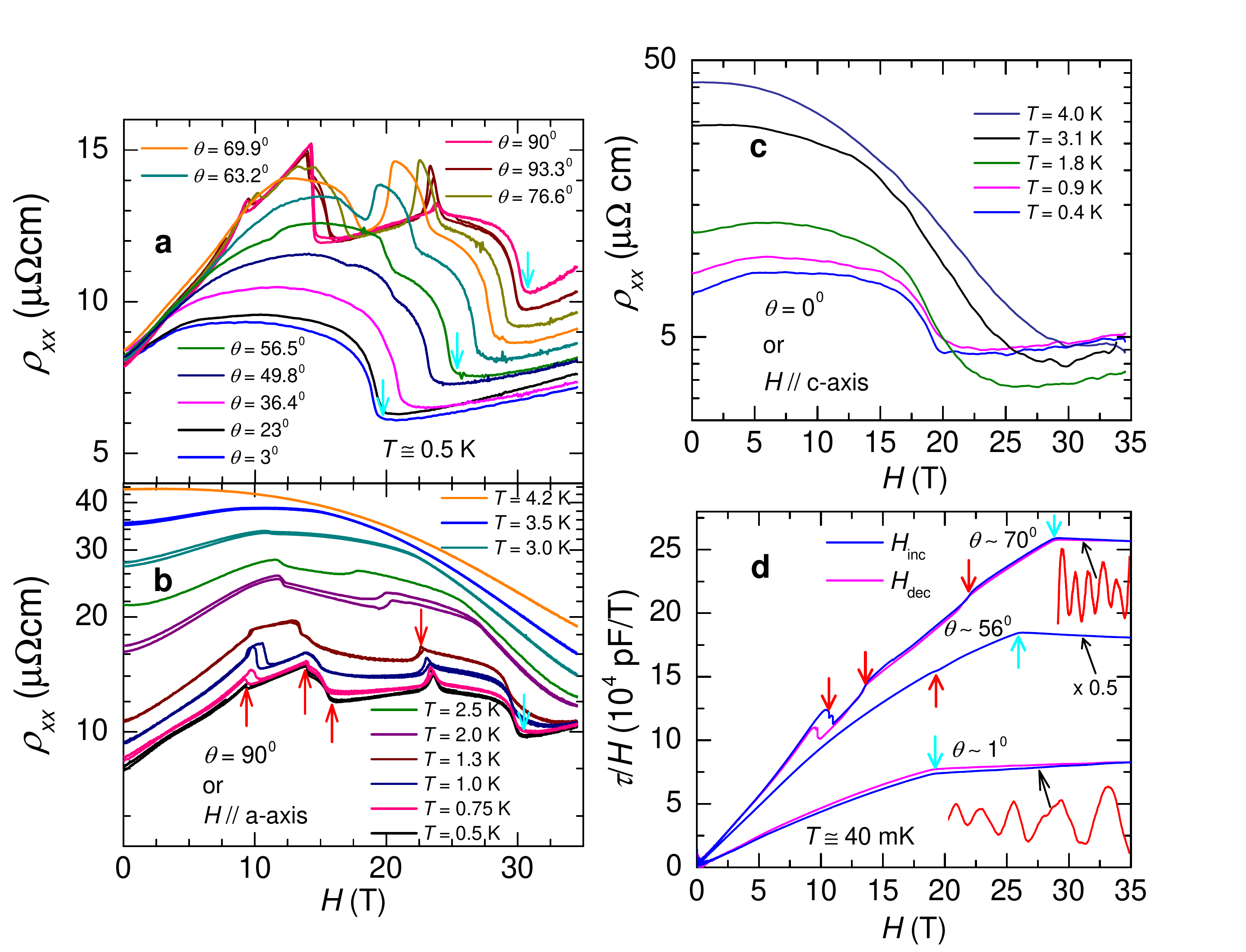}
\caption{
\textbf{a} Resistivity $\rho_{xx}$ as a function of $H$ for a CeCu$_2$Ge$_2$ single-crystal at a temperature $T = 0.5 $ K and for several angles $\theta$ between $H$ and the \emph{c}-axis. Notice that for $\theta = 3^{\circ}$ (or $H$ nearly parallel to the c-axis)  the positive magnetoresistance is followed by negative one, and subsequently, by a sharp change in slope and subsequent positive magnetoresistance indicating a phase-transition. Notice also that as $\theta$ increases additional structures are observable in $\rho_{xx}$ indicating additional phase transitions. \textbf{b} $\rho_{xx}$ as a function of $H$ and for several $T$s between 0.5 and 4.2 K. Just above the N\'{e}el transition, or at $T = 4.2$ K, $\rho_{xx}(H)$ decreases continuously as a function of $H$ reaching at $H = 34.5$ T less than 50 \% of its value at $H=0$ T. This indicates the field-induced suppression of spin-fluctuations. \textbf{c} Same as in \textbf{b} but for $\theta = 0^{\circ}$ or for $H \|$ c-axis. Notice i) the pronounced negative magnetoresistance at $T = 4.0$ K, i.e. nearly 100 \% between $H = 0$ and 35 T, and ii) the observation of a single anomaly at lower $T$s. \textbf{d} Magnetic torque $\tau$ normalized by $H$ and as a function of $H$ at $T = 40$ mK and for several angles $\theta$. Blue and magenta traces correspond to $H$-increasing and -decreasing sweeps, respectively. Notice that all the anomalies observed in the $\rho_{xx}(H)$ (indicated by red and cyan arrows) are associated with concomitant anomalies in $\tau(H,\theta)/H $ at the same critical-field values. For example, the sharp anomaly observed at $H \sim 20$ T for $(\theta = 0^{\circ})$, which defines the boundary between negative and positive magnetoresistive behavior at high fields and which is displaced to $\sim 30 $ T when $\theta = 90^{\circ}$, leads to a subsequent saturation in $\tau(H,\theta)/H$. This is a strong indication for the suppression of the SDW-phase observed at $H = 0$ T. The subtraction of a background unveils an oscillatory pattern in the paramagnetic (PM) region, or the de Haas van Alphen effect (red traces).}
\end{center}
\end{figure*}

Here, we show that the antiferromagnetic ground-state of CeCu$_2$Ge$_2$ can be continuously suppressed by an external magnetic field,
regardless of its orientation, leading to what seemingly is a 2$^{\text{nd}}$-order phase-transition line in the $H-\theta$ phase-diagram (where $\theta$ is the angle between $H$ and the \emph{c}-axis). This transition leads to the observation, through the de Haas-van-Alphen-effect (dHvA), of very small Fermi-surface pockets in the paramagnetic state, with areas ranging from $\sim 1$ to up to $\sim 1.4$ \% of the area of the first Brillouin-zone ($A_{\text{FBZ}}$), and which are characterized by light effective masses ranging from 2.5 to 4.5 $m_0$. This indicates the concomitant suppression of the Kondo-effect. For $\theta = 0^{\circ}$ we cannot detect any clear evidence for quantum-critical behavior around the metamagnetic critical-field $H_p^c \sim 19$ T, such as non-Fermi liquid behavior or severe effective mass renormalization. These effective masses increase by an order of magnitude as $H$ is rotated towards the \emph{ab}-plane, making these frequencies undetectable when $\theta > 56^{\circ}$. For $\theta = 90^{\circ}$, antiferromagnetism is suppressed through a series of metamagnetic transitions, revealing non-FL behavior at $H_p^{a} \sim 30$ T. Therefore, quantum-criticality in CeCu$_2$Ge$_2$ is tuned by two physical-parameters, field and angle, and remarkably, not necessarily triggered by the simultaneous suppression of the SDW order-parameter, Kondo-effect and coherence temperature ($T_{\text{coh}} \sim 5$ K).

\section{Methods}
Single crystals of CeCu$_2$Ge$_2$ were grown by a Cu-Ge self-flux method. 99.99\% pure Ce (AMES), 99.9999\% Cu (Furuuchi Chemical), and 99.999\% Ge (Rare Metallic Co.) were used as starting materials for the crystal growth. A starting composition of Ce$_{0.15}$Cu$_{1.00}$Ge$_{1.00}$ was arc-melted to produce an alloy button. This button was broken into small pieces and placed in an Al$_{2}$O$_{3}$ crucible, itself subsequently sealed in a quartz tube. The ampule was heated up to 1200 $^{\circ}$C and kept at this temperature for three hours. This ampule was cooled to 825 $^{\circ}$C at a rate of -4 $^{\circ}$C/h.  The excess flux was centrifuged at 825 $^{\circ}$C to obtain the resulting single crystals.
Samples were characterized by X-ray diffraction spectroscopy and back Laue method. Electrical transport measurements were performed through a conventional 4-terminal configuration using a Lock-In method. Torque magnetometry was performed using a capacitive cantilever beam method, whose capacitance was measured with an Andeen-Hagerling bridge. Measurements were performed either in a $^3$He cryostat, or in a dilution refrigerator, coupled to the resistive Bitter magnets of the NHMFL. Temperature was controlled by regulating either the $^3$He gas pressure of the $^3$He cryostat, or the temperature of the dilution fridge mixing chamber through the application of a constant heating power.

\section{Results and Discussion}
\begin{figure}[htb]
\begin{center}
\includegraphics[width = 7.6 cm]{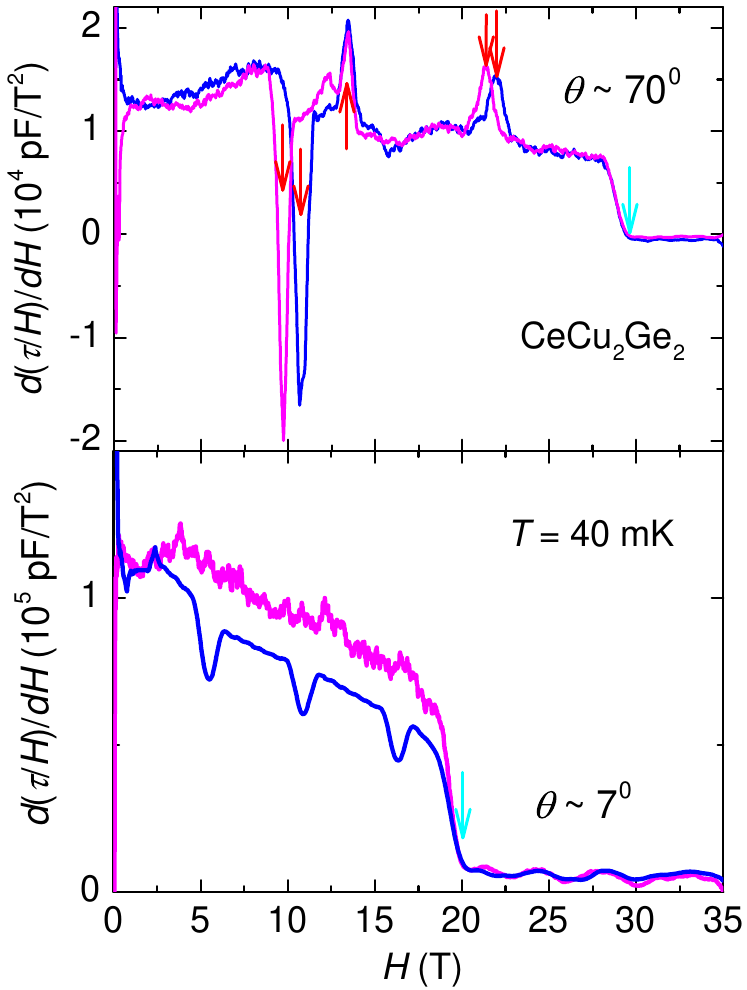}
\caption{
Derivative of the magnetic torque $\tau$ normalized by the field $H$ and as a function of $H$ for two angles, i.e. $\theta = 70^{\circ}$ and $7^{\circ}$, which are displayed in the top and bottom panels, respectively. Blue lines depict field-up sweeps while magenta lines depict the field-down ones. Red arrows indicate the position of the sharp metamagnetic transitions preceding the transition to the paramagnetic state which is indicated by cyan blue arrows. Notice the absence of hysteresis concerning the position in field for the transition towards the paramagnetic state. In the bottom panel the sharp dips in the blue trace are artifacts as discussed in the main text.}
\end{center}
\end{figure}
\begin{figure*}[htb]
\begin{center}
\includegraphics[width = 12 cm]{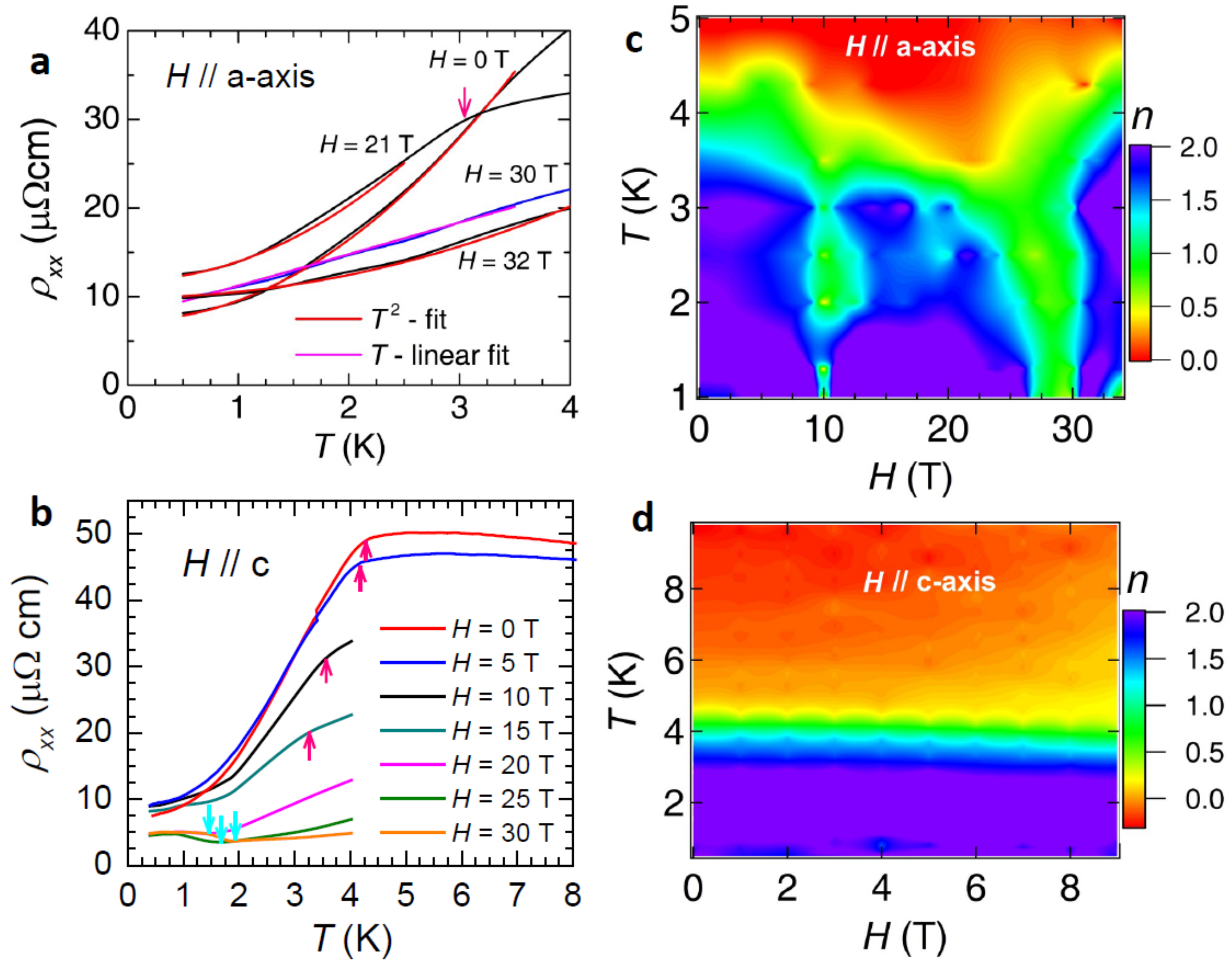}
\caption{
\textbf{a} $\rho_{xx}$ as a function of the temperature $T$, from the isothermal field scans in Fig. 1 \textbf{b}, and for several values of the field applied along the in-plane \emph{a}-axis. Red lines are fits to $\rho_{xx}(T) = \rho_0 + AT^2$, while the magenta line corresponds to a simple linear fit, suggesting non-FL like behavior for fields close to the critical field $H_p^a \simeq 30$ T required to suppress the SDW -state. \textbf{b} Same as in \textbf{a} but for $H$ applied along the \emph{a}-axis. In both Figs. \textbf{a} and \textbf{b}, the pink arrows indicate the onset of the phase-transition towards the SDW ground-state. Cyan arrows indicate the position in $T$ of the minima observed in $\rho_{xx}(T,H)$ when the PM-state is stabilized by $H$. Notice that the position in $T$ of these minima increases as the $H$ increases, indicating either a magnetoresistive effect or another phase-transition at lower $T$s. \textbf{c} Contour plot of the exponent $n = \partial \ln(\rho_{xx}(T)-\rho_0)/\partial \ln(T)$ built from a complete set of $\rho_{xx}(T,H)$ curves, such as the ones displayed in \textbf{a}. Notice the wide region in fields around $H_p^a$ where $\rho_{xx}(T)$ is  $\propto T$ indicating i) the suppression of the AF-order and ii) the emergence of non-FL behavior as expected for a putative QCP. The strongly hysteretic 1$^{\text{st}}$-order transition in the vicinity of $H = 10$ T, also leads to an anomalous $T$-dependence for $\rho_{xx}(T)$. \textbf{d} Same as in \textbf{c} but for $H$ along the \emph{c}-axis. Here, field values are limited to $H \lesssim 9 $ T, since above this field $\rho_{xx}(T)$ can no longer be described by a single power-law at low $T$s.}
\end{center}
\end{figure*}

Figure 1 \textbf{a} shows the in-plane resistivity $\rho_{xx}$ for a CeCu$_2$Ge$_2$ single-crystal as function of the field $H$ and for several angles $\theta$ between $H$ and the \emph{c}-axis of the crystal, at a temperature $T \simeq 0.5$ K. For $\theta = 0^{\circ}$ one observes positive magnetoresistance crossing-over towards a negative one until a pronounced change in slope (signaled by cyan arrows) is observed around $H_p^c \simeq 19$ T which is indicative of a phase-transition. As $H$ is rotated towards the \emph{ab}-plane, this feature continuously moves to higher fields, probably reflecting the anisotropy of the Land\'{e} \emph{g}-factor. Notice that for $\theta > 30^{\circ}$ additional structures (signaled by red arrows), i.e. new phase-transitions emerge in $\rho_{xx}$, becoming very pronounced as $\theta \rightarrow 90^{\circ}$. Fig. 1 \textbf{b} shows $\rho_{xx}(H, \theta = 90^{\circ})$ for several temperatures. For $T=4.2$ K, or just above the N\'{e}el transition, $\rho_{xx}(H, \theta = 90^{\circ})$ displays pronounced negative magnetoresistance decreasing by a factor of 2 when scanning the field from 0 to 35 T. This clearly indicates the field-induced suppression of pronounced spin-fluctuations and associated carrier scattering. As $T$ is reduced, one observes the progressive development of the previously shown series of anomalies in $\rho_{xx}(H, \theta = 90^{\circ})$. Fig. 1 \textbf{c} shows $\rho_{xx}(H, \theta = 0^{\circ})$ for a second CeCu$_2$Ge$_2$ single-crystal and for several values of temperature. At $T = 4.0$ K, $\rho_{xx}(H, \theta = 0^{\circ})$ decreases by nearly one-order of magnitude, or more precisely by $\sim 940$ \% between 0 and 35 T, which again can only be understood in terms of the field-induced suppression of pronounced spin-fluctuations. Notice however, that at low temperatures and above $H_p^c$, $\rho_{xx}(H, \theta = 0^{\circ})$ shows positive magnetoresistance indicating both the predominance of the orbital-effect and the near absence of scattering mediated by spin-fluctuations (the same can be said for fields along the \emph{ab}-plane when $H > H_p^{a}$). Therefore, at low temperatures pronounced spin fluctuations leading to negative magnetoresistance would seem to remain confined
to a narrow region in fields surrounding $H_p^{c}$, or $H_p^{a}$, as expected for putative QCPs. In order to evaluate the nature of the phase-transitions, i.e. 1$^{\text{st}}$- or 2$^{\text{nd}}$-order, we performed a thermodynamic measurement, or  magnetic torque $\mathbf{\tau} = \mathbf{M} \times \mathbf{H}$, where $\mathbf{M}$ is the sample's magnetization, as a function of $H$ at $T = 40$ mK. Fig. 1 \textbf{d} shows $\mathbf{\tau}(H)$ normalized by $H$, or $\sim \mathbf{M}(H)$ for 3 representative angles; increasing field-sweeps are depicted by blue lines, decreasing ones by magenta lines. Notice that i) for $\theta = 70^{\circ}$ several of the metamagnetic transitions occurring at $H < H_p$ are clearly hysteretic or 1$^{\text{st}}$-order in nature (Fig. 2 below shows the derivative of $\tau/H$ where the hysteresis becomes more apparent), ii) the saturation of $\tau/H$ above $H_p(\theta)$ indicates fully polarized moments or paramagnetism (PM), and iii) that for each value in $\theta$ the transition at $H_p(\theta)$ (indicated by cyan arrows) occurs exactly at the same field value for both field-up and down sweeps: at such low $T$s the absence of hysteresis is a strong indication for a 2$^{\text{nd}}$-order phase-transition at $H_p(\theta)$. As discussed below, the very small and apparent hysteresis observed in $\tau(H,\theta =1^{\circ})$ is an artifact produced by small variations in temperature (up to $\sim 15$ mK) associated with different field-up and -down sweep rates, which still lead to the exact same value for $H_p(\theta)$. Given that the excited CEF quasi-quartet is located at very high energies with respect to the ground-state doublet, it is quite unlikely that any of the observed metamagnetic transitions would be due to crossings among CEF levels. One is left with two scenarios i) ``local-moment" metamagnetism resulting from spin reorientations under $H$, and ii) field-induced SDW transitions (FISDW) resulting from a competition between the nesting vector and the inverse of the magnetic wave-length, as the area of the electronic orbit on the FS is required to accommodate an increasing number of magnetic flux quanta, and as observed in quasi-one-dimensional organic conductors.\cite{chaikin} The first scenario would imply a plausible progressive suppression of the Kondo-effect, but the resulting local-moment physics should lead to magnetization quasi-plateaus corresponding to stable spin-configurations within narrow field windows, in contrast to what is seen. The second scenario, only requires nearly nested two-dimensional FSs which have indeed been predicted by band structure calculations.\cite{zwicknagl} In addition, the FS reconstructed by the SDW transition is susceptible to the Zeeman effect, particularly in a strongly correlated material where the Land\`{e} \emph{g}-factor is likely to be larger than 2. The concomitant magnetic field-induced FS deformation would favor slightly distinct nesting-vectors at particular critical fields.
Therefore, CeCu$_2$Ge$_2$ could correspond to unique example of a moderately anisotropic material displaying a cascade of FISDW transitions.
Figure 1 \textbf{d} also displays the oscillatory component superimposed onto the $\tau(H)$ signal (red lines), or the dHvA-effect, obtained by subtracting a polynomial background. We were not able to detected the dHvA-effect within the N\'{e}el phase which is probably a consequence of much heavier effective masses, requiring even lower $T$s.
\begin{figure*}[htb]
\begin{center}
\includegraphics[width = 12 cm]{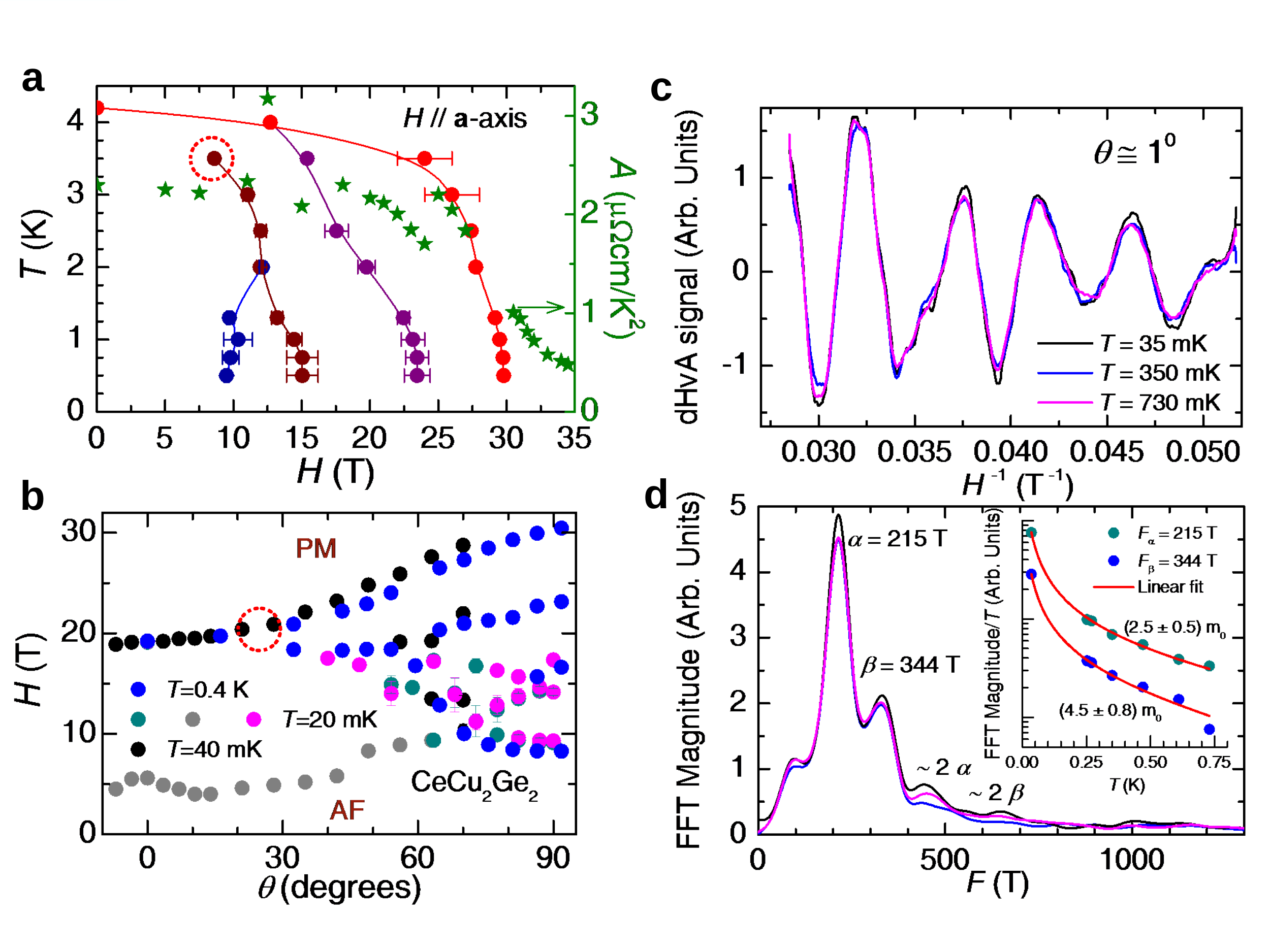}
\caption{
\textbf{a} $T$ as a function of $H\| a$-axis phase-diagram for CeCu$_2$Ge$_2$. AF and PM indicates the antiferromagnetic (SDW) and paramagnetic states, respectively. Red markers indicate the phase-boundary between the SDW and the PM-phase. Other lines indicate 1$^{\text{st}}$-order phase-transitions. Error bars depict the hysteresis associated with these transitions. The same graph shows the evolution of the $A = (\rho_{xx}(T) -\rho_0)/T^2$ coefficient; notice how it decreases when going from the SDW to the PM state. Red dotted line encircles the end-point of this 1$^{\text{st}}$-order line. \textbf{b} $H$ as a function of $\theta$ phase-diagram for CeCu$_2$Ge$_2$. Black markers were extracted from $\tau(\theta,H)$ at 40 mK, clear blue and grey markers from $\tau(\theta,H)$ at 20 mK, while blue and magenta markers were extracted from $\rho_{xx} (\theta, H)$ at 400 and 20 mK, respectively. Red dotted line encircles an area where a tricritical point should exist, i.e.  where a 1$^{\text{st}}$-order transition line meets a 2$^{\text{nd}}$-order one, and which should become a quantum tricritical point in the $T \rightarrow 0 \text{ K}$ limit. \textbf{c} Oscillatory component superimposed into the $\tau(H, \theta = 1^{\circ})$ in the PM-phase as a function of inverse field $H^{-1}$ and for several $T$s.  \textbf{d} Fast Fourier transform (FFT) of the oscillatory signal(s) shown in \textbf{c}, indicating two main peaks at $F_{\alpha} = 215$ T  and $F_{\beta} = 344$ T, respectively, and another possible frequency at $F \sim 100$ T. Inset: Magnitude of the main FFT peaks, normalized by $T$ and as a function of $T$. Red lines are fits to the Lifzhitz-Kosevich formula $A/\sinh X$ (where $X = 2 \pi^2k_B T/\hbar\omega_c = 14.69 \mu T/\overline{H}$, with $\mu$ being the effective mass and $\overline{H}$ the average field value), from which we extract the effective masses, $\mu_{\alpha} = (2.5\pm 0.5)m_0$ and $\mu_{\beta} = (4.5\pm 0.5)m_0$ respectively, where $m_0$ is the free electron mass.}
\end{center}
\end{figure*}
\begin{figure*}[htb]
\begin{center}
\includegraphics[width = 12 cm]{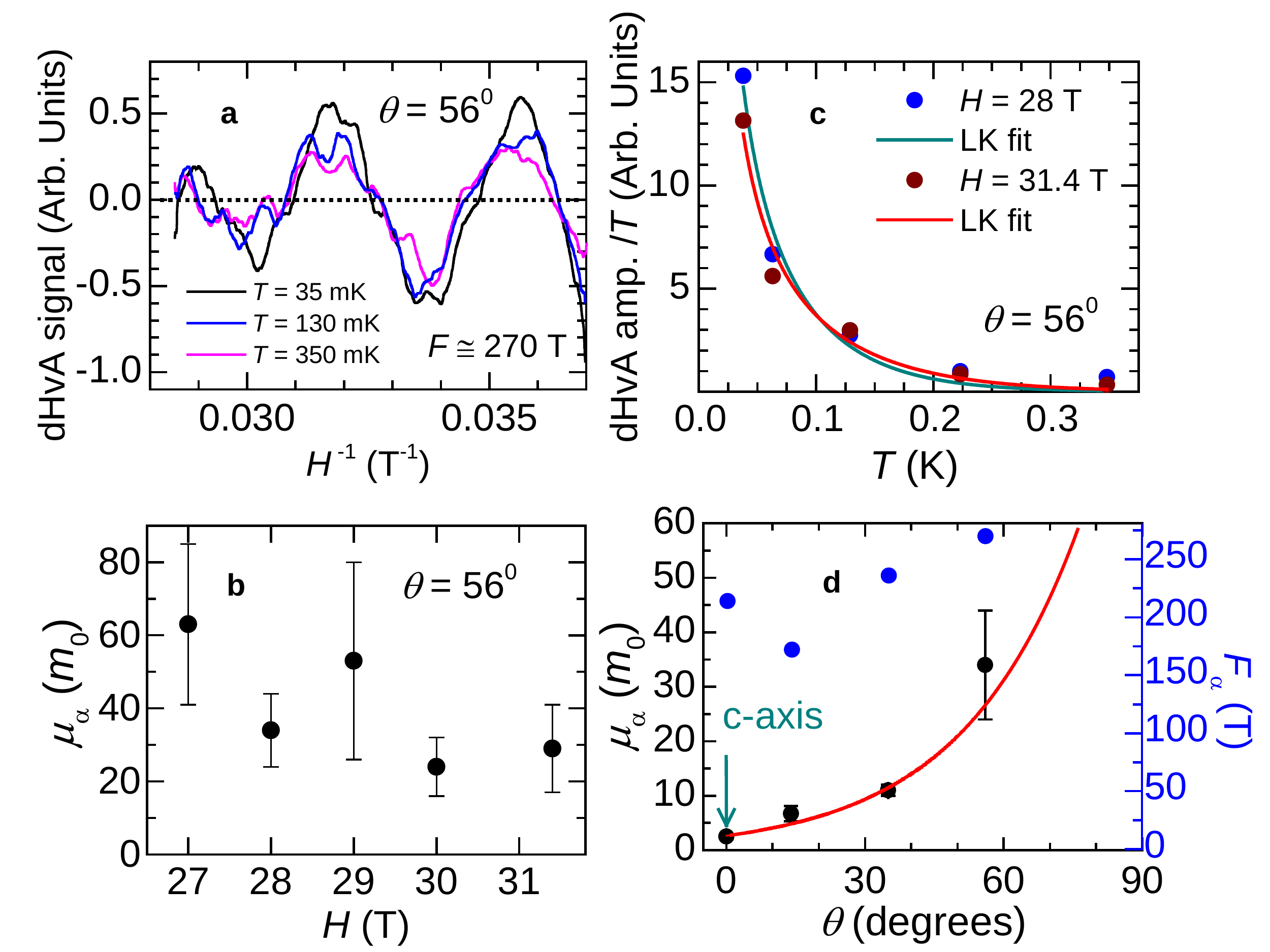}
\caption{
\textbf{a} Oscillatory component superimposed onto $\tau(H)$ measured within the paramagnetic phase of CeCu$_2$Ge$_2$ at an angle $\theta = 56^{\circ}$ and for several $T$s. At this angle the dHvA-signal is barely detectable, and in fact we could not clearly detect either $F_{\alpha}$ or $F_{\beta}$ at higher angles. Given the very few oscillations observed within this limited inverse field range, the frequency must be directly extracted from the oscillatory signal yielding $F \simeq 270$ T, or basically the average value between $F_{\alpha}(\theta \sim 0^{\circ})$ and $F_{\beta}(\theta \sim 0^{\circ})$. \textbf{b} Carrier effective mass, as extracted from the temperature dependence of the dHvA amplitude in \textbf{a} and for several field values. Notice that the extracted effective mass is nearly one order of magnitude higher than the masses extracted for $\theta = 0^{\circ}$. \textbf{c} Two examples of the effective mass extraction used to build Fig. \textbf{b}, i.e. Lifshitz-Kosevich fits of the amplitude of the dHvA signal as a function of $T$ for two field values. \textbf{d} $\mu_{\alpha}$ as a function of $\theta$. Red line is a fit to a single exponential suggesting an exponential divergence of $\mu_{\alpha}$ as $\theta$ approaches $90^{\circ}$. The same figure also shows $F_{\alpha}$ as a function of $\theta$, where the point at $\theta = 56^{\circ}$ was extracted from \textbf{a}.}
\end{center}
\end{figure*}

Figure 2 displays the derivative of the magnetic torque $\tau$ normalized by the field $H$ with respect to $H$ and as function of $H$ at $T \simeq 40$ mK, to unveil the hysteresis or the absence of thereof, associated with some of the metamagnetic transitions. The term ``metamagnetic” is used here to indicate a super-linear increase in $\tau/H \propto M$ where $M$ is the magnetization. Blue lines depict data collected during field-up sweeps, while the magenta lines correspond to the ones collected during field-down sweeps. The metamagnetic transitions preceding the suppression of the SDW-state are indicated by red arrows, while the transition towards the paramagnetic state is indicated by cyan blue arrows. Notice that the transitions indicated by the red arrows, lead to sharp features in $\partial(\tau/H)/\partial H$ and in most cases, also to a pronounced hysteresis which are indications for 1$^{\text{st}}$-order phase-transitions. Most importantly, for both orientations the transition towards the paramagnetic state is clearly non-hysteretic, even at these low temperatures, indicating they are 2$^{\text{nd}}$-order in nature. In the bottom panel, the anomalies or dips in the field-up sweep trace (blue) are artifacts associated with the field-up sweep pattern: $H$ was increased at a higher rate than the one used for the field-decreasing sweep trace. This higher rate induces eddy currents in the probe and in the cryostat leading to an increase in the temperature of the dilution refrigerator, i.e. by about $\sim 15$ to $20$ mK. Each dip corresponds to moments in time where the sweep rate was brought down to zero, leading to a momentary decrease in temperature. Therefore, the apparent hysteresis between field-up and -down traces can be simply attributed to the extreme sensitivity of the torque technique to small changes in $T$ induced by the distinct field sweep rates. We have expressly chosen both traces to illustrate that even with these less than ideal circumstances, one does not detect any clear hysteresis in the position in field where the phase-transition towards the paramagnetic state occurs. This is a clear indication for a 2$^{\text{nd}}$-order phase-transition, which ought to become quantum-critical in $T \rightarrow 0$ K limit.

Figure 3 \textbf{a} displays the in-plane resistivity $\rho_{xx}$ as a function of $T$, as extracted from the isothermal field-scans in Fig. 1, for a few field values applied along the \emph{a}-axis. As seen, at zero or low fields and at the lowest $T$s, $\rho_{xx}(T)-\rho_0 \propto T^2$ but for fields around $H_p^{a} \sim 30$ T $\rho_{xx}(T)$ becomes linear in $T$ therefore displaying the characteristic non-Fermi liquid behavior observed around QCPs resulting from the suppression of a SDW. \cite{custers} Notice that the $\Delta \rho_{xx} \propto T^2$, or conventional FL-behavior, is recovered at higher fields (see 32 T trace), indicating that one has driven the system away from the $H-T$ region dominated by the quantum-critical fluctuations. In contrast, as shown in Fig. 3 \textbf{b} for fields along the \emph{c}-axis, $\rho_{xx}(T)$ evolves from the $T^2$ dependence towards a $T-$dependence which cannot be described by a single power law: a minimum in $\rho_{xx}(T)$ emerges (cyan arrows) and moves to higher $T$s as $H$ increases, suggesting either the emergence of another phase, concealing a possible QCP, or a simple magnetoresistive effect. On the other hand, in the N\'{e}el state the evolution of the $A = (\rho_{xx}(T)-\rho_0)/T^2$ coefficient as a function of $H \|c$-axis, does not provide any clear indication for the proximity to a QCP, which is typically a power law dependence; $A\propto(H-H_p)^{\alpha}$ with $\alpha \sim -1$.\cite{custers} $A$ just decreases slightly as $H$ increases. To illustrate the evolution of the exponent $n = \partial\ln(\rho_{xx}-\rho_0)/\partial\ln T$ as a function of both $T$ and $H$, we generated contour plots from a number of $\rho_{xx}(T)$ traces acquired under fixed fields applied either along the \emph{a}- or the \emph{c}-axis shown in Figs. 3 \textbf{c} and \textbf{d}, respectively. As seen in Fig. 3 \textbf{c}, at $H = 0 $ T FL behavior or $n \simeq 2$ emerges below $\sim 2$ K, but there are two field regions, centred around $\sim 10$ and $\sim 29.5$ T respectively, where one clearly sees that $n \simeq 1$. The first region according to Figs. 1 and 2, corresponds to a strongly hysteretic phase-transition which was previously claimed to be associated with a possible quantum-critical point for $H\| [110]$-direction where it takes place around $H_c\sim 8$ T.\cite{dksingh} As it is discussed below, this transition for $H\| [100]$-direction corresponds to a $1^{\text{st}}$-order line in the $T-H$ diagram which at higher $T$s ends at a well-defined end-point, where it becomes strictly $2^{\text{nd}}$-order in nature.\cite{grigera} As is also discussed below, this transition \emph{disappears} as $H$ is tilted towards the \emph{c}-axis, implying that this end-point must move to $T= 0 \text{ K}$ at a certain angle, where it would become a quantum-critical end-point. Thus, at low-$T$s the coexistence of non-FL behavior likely due to QC-fluctuations, with a strongly $1^{\text{st}}$-order transition, would be an indication for its proximity to a quantum-critical end-point (accessible upon field rotation), and this would reconcile our observations with those of Ref. \onlinecite{dksingh}.

The second region centred around $H_p^a \sim 29.5$ T is where $\tau/H$ reaches saturation or a PM-state due to the suppression of the SDW-state, with the anomalous exponent indicating concomitant quantum-critical fluctuations. For fields along the \emph{c}-axis on the other hand, one is forced to limit $H$ to values $< 10$ T to extract a well-defined $n$ since above this field value, and within $ T \simeq 0.4$ and $4$ K, $\rho_{xx}(T)$ cannot be described by a single power law.

Figures 4 \textbf{a} and 4 \textbf{b} show the resulting phase-diagrams built from the torque and the transport measurements, or the transition-temperatures as a functions of $H$ applied along the \emph{a}-axis, and the transition fields as functions of $\theta$, respectively. The discontinuity in the heat-capacity,\cite{fisher} as well as our transport and torque measurements, indicate that the red line in Fig. 4 \textbf{a}, which defines the boundary between PM and N\'{e}el states corresponds to a 2$^{\text{nd}}$-order phase transition. All the other lines correspond to either strong or weak 1$^{\text{st}}$-order transitions where the error bars indicate the hysteresis in field.
The same Fig. 4 \textbf{a} shows the $A$-coefficient as a function of $H$ (green markers), which is proportional to the quasiparticle-quasiparticle scattering cross-section. As seen, the $A$-coefficient decreases by a factor of nearly $\sim 5$ when going from the SDW to the PM state. Since at $H=0$, the $A$-coefficient is related to the electronic contribution to the specific heat $\gamma_0$ through the Kadowaki-Woods ratio $R_{\text{KW}}=A/\gamma_0^2$ such a decrease in $A$ points towards a severe reduction in the density of states at the Fermi-level or equivalently, to a strong decrease in quasiparticle effective-masses and electron-electron interactions at the SDW to PM transition. In Fig. 4 \textbf{b} according to our experimental evidence, the phase-boundary between AF and PM states is 2$^{\text{nd}}$-order in nature suggesting the possibility of a quantum-critical line in the $H-\theta$ diagram as well as the existence of a \emph{quantum-tricritical point} (in $T \rightarrow 0$ K limit) at the angle where it bifurcates.\cite{misawa,jakubczyk1,jakubczyk2} All the other lines correspond to 1$^{\text{st}}$-order transitions, except for the gray line which results from a very mild, barely detectable anomaly in $\tau/H$. Notice that all these 1$^{\text{st}}$-order lines disappear as $H$ is rotated towards the \emph{c}-axis. This suggests that 1$^{\text{st}}$-order transition end points, as for example the end of the brown line in Fig. 4 \textbf{a} which should be a 2$^{\text{nd}}$-order phase transition point, might become a quantum-critical end point (QCEP) in the $T \rightarrow 0 $ K limit as seen in Sr$_3$Ru$_2$O$_7$,\cite{grigera} and as $H$ is rotated towards the \emph{c}-axis. In fact, the relative proximity to a QCEP upon rotation might perhaps explain the anomalous exponent in the $\rho(T)$ around the strong 1$^{\text{st}}$-order transition observed at $H \sim 10$ T.

Figure 4 \textbf{c} displays the dHvA signal superimposed onto $\tau/H$ for an angle $\theta = 1^{\circ}$ and for several $T$s. As seen the dHvA amplitude shows a small $T$-dependence between 35 and 730 mK indicating relatively light effective-masses. An analysis of the dHvA amplitude as a function of field, does \emph{not} unveil mass renormalization upon approaching $H_p^c$, as one would expect for the presence of quantum critical fluctuations. The fast Fourier transform of the oscillatory signal is displayed in Fig. 4 \textbf{d}, which reveals two main peaks at $F_{\alpha}= 215$ T (corresponding to a cross-sectional area of $\sim 1$ \% of the $A_{\text{FBZ}}$) and  $F_{\beta}=344$ T ($\sim 1.4$ \% of the $A_{\text{FBZ}}$), with corresponding effective masses $\mu_{\alpha}= (2.5 \pm 0.5)$ and $\mu_{\beta}= (4.5 \pm 0.8)$ $m_0$, as extracted from the Lifshitz-Kosevich fits shown in the inset of Fig. 4 \textbf{d}. We could not detect these or other frequencies for $H < H_p^a$, suggesting considerably heavier effective masses in the N\'{e}el state. Relatively light masses are in sharp contrast with those extracted from optical conductivity \cite{armitage} or with a $\gamma_0$-coefficient approaching 200 mJ/K$^2$mol.\cite{fisher} Coupled to the sharp decrease of the $A$-coefficient at $H_p^a$, light masses in the PM-state would point to the suppression of the Kondo-effect at $H_p(\theta)$.

Since SDW related QCPs are claimed to lead to hot spots on the Fermi surface associated with the nesting vector,\cite{gegenwart} we proceed to explore the $\theta$ dependence of the dHvA effective masses. Both $\mu_{\alpha}$ and $\mu_{\beta}$ are observed to increase considerably as $\theta$ increases towards the \emph{a}-axis, or as the component of electronic orbit perpendicular to $H$ explores the inter-planar direction, becoming undetectable for $\theta \gtrsim 56^{\circ}$. As seen in Fig. 5 \textbf{a}, for $\theta = 56^{\circ}$ the dHvA signal is very small (therefore noisier) showing a more pronounced $T$-dependence when compared to $\theta=1^{\circ}$. The envelope of the dHvA signal clearly indicates beating between close frequencies, such as $F_{\alpha}$ and $F_{\beta}$, but the limited number of oscillations prevents an accurate FFT extraction of the corresponding $F$s. A direct reading yields $F = 270$ T, or an average value between  $F_{\alpha}$ and $F_{\beta}$. The extraction of the corresponding effective mass $\mu$, from the amplitude of the dHvA signal as a function of $T$, and for several $H$ values is shown in Fig. 5 \textbf{b}. The extracted $\mu$ values are one order of magnitude larger than those extracted for $\theta = 1^{\circ}$ suggesting also an increase in $\mu$ as $H \rightarrow H_p$, although the error bars are too large to reach a definitive conclusion. Representative Lifshitz-Kosevich fits are shown in Fig. 5 \textbf{c}. Finally, Fig. 5 \textbf{d} shows both $\mu_{\alpha}$ and $F_{\alpha}$ as a function of $\theta$, revealing a severe renormalization of $\mu_{\alpha}$ (red line is an exponential fit) as $\theta \rightarrow 90^{\circ}$,  although the values for $\theta = 56^{\circ}$ correspond to averages between the
$\alpha$ and $\beta$ orbits.

\section{Summary/Conclusions}

In summary, we find that the Kondo and coherence temperatures as well as the spin-density wave transition-temperature can be suppressed by the application of an external magnetic field through a second-order phase-transition, regardless of its crystallographic orientation. A priori this would suggest a few possibilities: i) A 2$^{\text{nd}}$-order line in the $H-\theta$ phase diagram, leading to a quantum-critical line where all relevant energy scales, N\'{e}el, coherence and Kondo-temperatures collapse leading to a line of quantum critical points, ii) quantum-criticality driven solely by the suppression of the Kondo-effect, which is consistent with the light masses observed in the PM-state for $\theta \sim 1^{\circ}$, and iii) QC driven solely by the suppression of SDW-state leading to hot spots on the Fermi surface associated with the nesting wave-vector, and which would lead to heavy electron-masses. Since we could not detect heavy masses, or non-FL behavior for fields along the \emph{c}-axis or in wide angular window around it, our results are difficult to reconcile with scenarios i) and ii). However, according to prevailing quantum-critical scenarios a 2$^{\text{nd}}$-order line in the field as a function of angle phase-diagram, where all relevant energy scales, in particular the spin-density wave transition-temperature, are continuously suppressed by the magnetic-field, should lead to a line of quantum critical points in the limit of zero temperature. Therefore, our study unveiling non-Fermi liquid behavior and mass divergence only when the field is aligned along the $a$-axis is a clear challenge to our current understanding of quantum criticality (in contrast to recent theoretical progress in describing model a system \cite{imai}) and suggests the existence of Fermi surface hot spots located nearly along the $k_z$ direction most likely associated with a three dimensional nesting wave-vector. Hence, in the limit of very low temperatures the observation of a second-order phase transition associated to the suppression of an order-parameter by tuning a physical parameter, such as magnetic field, would seem not to be a sufficient condition for the observation of quantum criticality.

We want to finish by emphasizing that the complexity of the CeCu$_2$Ge$_2$ phase-diagram as a function of temperature, field and angle, which suggests the possible existence of quantum tricritical and quantum-critical end points associated with multiple 1$^{\text{st}}$-order metamagnetic transitions, will contribute to enrich our overall understanding of quantum criticality. For example, the results in Ref. \onlinecite{dksingh} indicate a quantum phase-transition for $H \simeq 8$ T applied along the [-110]-direction, which is observed by us to become strongly 1$^{\text{st}}$-order (albeit coexisting with non-Fermi liquid behavior) when $H \simeq 10$ T is applied along the [100]-direction. This suggests the existence of a QC end-point which can be accessed by varying the field-orientation with respect to the crystallographic axes. However, it remains unclear at the moment, from ours as well as the results in Ref. \onlinecite{dksingh}, if such a QC end-point would be directly accessed by varying temperature and angle or if phase-formation would conceal it.\cite{grigera2,yelland} This aspect, as well as a full comparison between the overall physical properties observed at low temperatures around $H \| [-110] \simeq 8 \text{ T}$ and $H \| [100] \simeq 29.5 \text{ T}$ deserve future attention. As for quantum tri-critical point(s) it is a concept claimed to be relevant for the high-temperature superconductivity observed in the Fe arsenide superconductors\cite{giovannetti} or for the anomalous non-Fermi liquid behavior observed in MnSi.\cite{pfleiderer} Insofar it has been experimentally challenging to precisely tune either system towards a putative quantum tri-critical point in order to expose it and in this way confirm its conceptual relevance. In contrast, through this study we learned that we can in CeCu$_2$Ge$_2$ precisely access a tri-critical point by tuning two non-thermal parameters namely field and angle at very low temperatures. Hopefully, such a tunability might allow us to unambiguously unveil a quantum tri-critical point.

\section{Acknowledgements}
The NHMFL is supported by NSF through NSF-DMR-1157490 and the
State of Florida. YS acknowledges support from ICAM. 
PS Is partially supported by DOE-BES through grant DE-FG02-98ER45707.
LB is supported by DOE-BES through award DE-SC0002613.

\end{document}